\documentclass{SciPost}
\usepackage[bitstream-charter]{mathdesign}

\hypersetup{
    colorlinks,
    linkcolor={red!50!black},
    citecolor={blue!50!black},
    urlcolor={blue!80!black}
}
\DeclareSymbolFont{usualmathcal}{OMS}{cmsy}{m}{n}
\DeclareSymbolFontAlphabet{\mathcal}{usualmathcal}
\fancypagestyle{SPstyle}{
\fancyhf{}
\lhead{\colorbox{scipostblue}{\bf \color{white} ~SciPost Physics Core }}
\rhead{{\bf \color{scipostdeepblue} ~Submission }}

\fancyfoot[C]{\textbf{\thepage}}
}

\begin{document}

\pagestyle{SPstyle}

\begin{center}
{\Large \textbf{%
\color{scipostdeepblue}{
Coalescing hardcore-boson condensate states with nonzero momentum\\
}}}

\textbf{\ C. H. Zhang, Z. Song\textsuperscript{$\dagger$} }

\textbf{\ }School of Physics, Nankai University, Tianjin 300071, China 
%%%%%%%%%% END TODO: AFFILIATIONS
% Provide email address of corresponding author(s)
\\[0pt]
$\dagger $ \href{mailto:songtc@nankai.edu.cn}{{\small songtc@nankai.edu.cn}} 
\end{center}

\section*{\color{scipostdeepblue}{Abstract}}

\textbf{Exceptional points (EPs), as an exclusive feature of a non-Hermitian
system, support coalescing states to be alternative stable state beyond the
ground state. In this work, we explore the influence of non-Hermitian
impurities on the dynamic formation of condensate states in one-, two-, and
three-dimensional extended Bose-Hubbard systems with strong on-site
interaction. Based on the solution for the hardcore limit, we show exactly
that condensate modes with off-diagonal long-range order (ODLRO) can exist
when certain system parameters satisfy specific matching conditions. Under
open boundary conditions, the condensate states become coalescing states
when the non-Hermitian $\mathcal{PT}$-symmetric boundary gives rise to the
EPs. The fundamental mechanism behind this phenomenon is uncovered through
analyzing the scattering dynamics of many-particle wavepackets at the
non-Hermitian boundaries. The EP dynamics facilitate the dynamic generation
of condensate states with non-zero momentum. To further substantiate the
theoretical findings, numerical simulations are conducted. This study not
only unveils the potential condensation of interacting bosons but also
offers an approach for the engineering of condensate states. }

\vspace{\baselineskip}

%%%%%%%%%% BLOCK: Copyright information
% This block will be filled during the proof stage, and finilized just before publication.
% It exists here only as a placeholder, and should not be modified by authors.
\noindent%
\textcolor{white!90!black}{\fbox{\parbox{0.975\linewidth}{\textcolor{white!40!black}{\begin{tabular}{lr}  \begin{minipage}{0.6\textwidth}    {\small Copyright attribution to authors. \newline
    This work is a submission to SciPost Physics Core. \newline
    License information to appear upon publication. \newline
    Publication information to appear upon publication.}
  \end{minipage} & \begin{minipage}{0.4\textwidth}
    {\small Received Date \newline Accepted Date \newline Published Date}  \end{minipage}
\end{tabular}}
}}
} %%%%%%%%%% BLOCK: Copyright information

% For convenience during refereeing we turn on line numbers:


\vspace{10pt} \noindent\rule{\textwidth}{1pt} 
\tableofcontents
\noindent\rule{\textwidth}{1pt}
\vspace{10pt}

% Write your article contents here, starting from first \section.
% An example structure is given below.

\section{Introduction}

Recent developments in cold atom experiments provide a versatile platform
for realizing various phases of interacting and non-interacting bosonic
systems \cite%
{bloch2012quantum,atala2014observation,aidelsburger2015measuring,stuhl2015visualizing}
. Available experimental setups nowadays allow for the control of both
geometry and interactions, so as to investigate the real-time evolution of
quantum many-body systems directly with engineered model Hamiltonians \cite%
{jane2003simulation,bloch2012quantum,blatt2012quantum}. It thus boosts the
theoretical predictions of exotic quantum phases in interacting systems,
which then might be realized and tested in experiments. Exact solutions for
quantum many-body systems are rare, but important for providing valuable
insights for the characterization of new forms of quantum matter and dynamic
behaviors.

Bose-Einstein condensation (BEC) is one of the most striking manifestations
of the quantum nature of matter on the macroscopic scale \cite%
{bose1924plancks}. It represents a formation of a collective quantum state
of free bosons. Intuitively, on-site repulsive interactions should block the
formation of BEC under the moderate particle density. A lot of effort has
been devoted to investigate and understand the role of particle-particle
interactions on the occurrence of BEC \cite{shi1998finite,andersen2004theory}%
.

On the other hand, recent years have seen a growing interest in
non-Hermitian descriptions of condensed-matter systems \cite%
{Lee2016,Kunst2018,Yao2018,Gong2018,El-Ganainy2018,Nakagawa2018,Shen2018,Wu2019,Yamamoto2019,Song2019,Yang2019,Hamazaki2019,Kawabata2019,Kawabata2019a,Lee2019,Yokomizo2019,Jin2020}.
It has been shown that the interplay between non-Hermiticity and interaction
can give rise to exotic quantum many-body effect, ranging from non-Hermitian
extensions of Kondo effect \cite%
{Nakagawa2018,Lourenifmmodemboxccelseccfio2018}, many-body localization \cite%
{Hamazaki2019}, Fermi surface in coordinate space \cite{Mu2020}, to
fermionic superfluidity \cite{Yamamoto2019,Okuma2019}. The cooperation
between the non-Hermiticity and interaction may lead to rich quantum phases
due to the peculiarity of non-Hermitian system.

Exceptional points (EPs), as an exclusive feature of a non-Hermitian system,
are degeneracies of non-Hermitian operators \cite%
{Berry2004,Heiss2012,Miri2019,Zhang2020}. The corresponding eigenstates
coalesce into one state, resulting in the incompleteness of Hilbert space.
The peculiar features around EP have sparked tremendous attention to the
classical and quantum photonic systems \cite%
{Doppler2016,Xu2016,Assawaworrarit2017,Wiersig2014,Wiersig2016,Hodaei2017,Chen2017}%
.Notably, a coalescing state has an exclusive feature. On the one hand, it
is an eigenstate of the Hamiltonian, on the other hand, it has the advantage
that it is also a target state for a long-time evolution of various initial
states. In this sense, a coalescing state is an alternative stable state
beyond the ground state. Given the above rapidly growing fields in
experimental and theoretical perspectives, we are motivated to investigate
the impact of non-Hermitian impurities on the dynamic formation of
condensate states of interacting bosons.

In this paper, we study one-, two-, and three-dimensional extended
Bose-Hubbard systems with strong on-site interaction. The exact solution for
the hardcore limit shows that there exists condensate modes when the system
parameters meet the matching conditions. It also allows us to calculate the
correlation function for any size system, so as to prove that condensate
states\ indeed possesses off-diagonal long-range order (ODLRO) \cite%
{yang1962concept}. We focus on the impact of non-Hermitian impurities on the
dynamic formation of condensate states. For open boundary condition, the
condensate states become coalescing states when the non-Hermitian $\mathcal{%
PT}$-symmetric boundary induces the EPs. The underlying mechanism is
revealed by the reflectionless absorption of many-particle wavepacket with
resonant momentum by the non-Hermitian boundary. In parallel, the EP
dynamics allows the dynamic generation of condensate states with nonzero
momentum. We perform numerical simulations for finite-size system\ to
demonstrate and verify the theoretical results. The finding not only reveals
the possible condensation of interaction bosons, but also provides a method
for condensate state engineering in an alternative way. The implications of
this work are significant for both theoretical and practical applications in
the realm of quantum many-body systems and could pave the way for innovative
strategies in quantum state manipulation and control.

This paper is organized as follows. In Sec. \ref{Model and condensate states}%
, we introduce the model Hamiltonian and its condensate eigenstate. In Sec. %
\ref{Coalescing condensate states}, we derive the exact condition for
coalescing condensate state. In Sec. \ref{Resonant scattering of
non-Hermitian impurity}, we perform the numerical simulations to demonstrate
reflectionless scattering of many-particle Gaussian wavepacket at
non-Hermitian boundary. In Sec. \ref{Dynamic generation of condensate states}%
, we present the possibility of dynamically generating condensate state with
an arbitrary initial state. Finally, we summarize our results in Sec. \ref%
{Summary}.

\section{Model and condensate states}

\label{Model and condensate states}

We start our study from a general form of the Hamiltonian on a
three-dimensional lattice $N_{1}\times N_{2}\times N_{3}$ 
\begin{eqnarray}
H =\sum\limits_{\alpha =1}^{3}J_{\alpha }\sum_{\mathbf{r}}\frac{1}{2}\hat{a%
}_{\mathbf{\ r}}^{\dagger }\hat{a}_{\mathbf{r}+\mathbf{e}_{\alpha }}+\mathrm{%
H.c.} +\sum\limits_{\alpha =1}^{3}V_{\alpha }\sum_{\mathbf{r}}\hat{n}_{%
\mathbf{r}} \hat{n}_{\mathbf{r}+\mathbf{e}_{\alpha }} +\sum\limits_{\alpha =1}^{3}\sum_{\mathbf{r}}\left( \mu _{\alpha }\hat{n}%
_{ \mathbf{r}}\delta _{1,m_{\alpha }}+\mu _{\alpha }^{\ast }\hat{n}_{\mathbf{%
r} }\delta _{N_{\alpha },m_{\alpha }}\right),  \label{H}
\end{eqnarray}%
where $\hat{a}_{\mathbf{r}}^{\dagger }$ is the hardcore boson creation
operator at the position $\mathbf{r=}m_{1}\mathbf{e}_{1}+m_{2}\mathbf{e}%
_{2}+m_{3}\mathbf{e}_{3}$ $(m_{\alpha }=1,2,...,N_{\alpha },$ $\alpha
=1,2,3) $, satisfying%
\begin{equation}
\left\{ \hat{a}_{l},\hat{a}_{l}^{\dagger }\right\} =1,\left\{ \hat{a}_{l}, 
\hat{a}_{l}\right\} =0,
\end{equation}%
and%
\begin{equation}
\left[ \hat{a}_{j},\hat{a}_{l}^{\dagger }\right] =0,\left[ \hat{a}_{j},\hat{%
a }_{l}\right] =0,
\end{equation}%
for $j\neq l$, and $\hat{n}_{\mathbf{r}}=\hat{a}_{\mathbf{r}}^{\dagger }\hat{%
	a}_{\mathbf{r}}$, $\mathbf{e}_{\alpha }$ is the unit vector for $N_{\alpha
} $. Under the open boundary condition, we define $\hat{a}_{\mathbf{r+}
N_{\alpha }\mathbf{e}_{\alpha }}=0$, while $\hat{a}_{\mathbf{r+}N_{\alpha } 
\mathbf{e}_{\alpha }}=\hat{a}_{\mathbf{r}}$ for the periodic boundary
condition ($\alpha =1,2,3$).

The parameters are taken as%
\begin{equation}
\left\{ 
\begin{array}{c}
V_{\alpha }=J_{\alpha }\cos q_{\alpha } \\ 
\mu _{\alpha }=J_{\alpha }\frac{e^{iq_{\alpha }}}{2}%
\end{array}
\right. ,  \label{resonant condition}
\end{equation}%
with arbitrary real number $q_{\alpha }$\ for the case with open boundary
condition, but with $\mathbf{q}=(q_{1},q_{2},q_{3})$, $q_{\alpha }=2\pi
m_{\alpha }/N_{\alpha }$ $(m_{\alpha }=1,2,...,N_{\alpha },$ $\alpha =1,2,3)$
and $\mu _{\alpha }=0$\ with periodic boundary condition. When taking $%
\left\{ N_{\alpha }\right\} =\left( N_{1},N_{2},N_{3}\right) =\left(
N_{1},N_{2},1\right) $ or\ $\left( N_{1},1,1\right) $, the system reduces to
two- or one-dimensional systems.

In the following, we will show that state

\begin{eqnarray}
\left\vert \psi _{n}\right\rangle &=&\frac{1}{\Omega _{n}}\left( \sum_{%
\mathbf{r}}\hat{a}_{\mathbf{r}}^{\dagger }e^{-i\mathbf{q\cdot r}}\right)
^{n}\left\vert 0\right\rangle , \\
\Omega _{n} &=&\frac{1}{(n!)\sqrt{C_{N}^{n}}},
\end{eqnarray}%
is an eigenstate of the system, where the vacuum state $\left\vert
0\right\rangle =\prod_{\mathbf{r}}\left\vert 0\right\rangle _{\mathbf{r}}$,
with\ $\hat{a}_{\mathbf{r}}\left\vert 0\right\rangle _{\mathbf{r}}=0$. In
both two cases (also including mixed boundary conditions), the Hamiltonian
can be written as the form%
\begin{equation}
H=\sum\limits_{\alpha =1}^{3}\sum_{\mathbf{r}}h_{\mathbf{r}}^{\alpha
}+\sum\limits_{\alpha =1}^{3}V_{\alpha }\hat{n},
\end{equation}%
where the dimer term is non-Hermitian, i.e.,%
\begin{eqnarray}
h_{\mathbf{r}}^{\alpha }=J_{\alpha }\left[\frac{1}{2}\hat{a}_{\mathbf{r}%
}^{\dagger }\hat{a}_{\mathbf{r}+\mathbf{e}_{\alpha }}+\mathrm{H.c.}+\cos (%
\mathbf{q\cdot e}_{\alpha })(\hat{n}_{\mathbf{r}}\hat{n}_{\mathbf{r}+\mathbf{%
e}_{\alpha }}-\hat{n}_{\mathbf{r}}-\hat{n}_{\mathbf{r}+\mathbf{e}_{\alpha }})
+\frac{1}{2}\left( e^{i\mathbf{q\cdot e}_{\alpha }}\hat{n}_{\mathbf{r}%
}+e^{-i\mathbf{q\cdot e}_{\alpha }}\hat{n}_{\mathbf{r}+\mathbf{e}_{\alpha
}}\right) \right],
\end{eqnarray}%
and $\hat{n}=\sum_{\mathbf{r}}\hat{n}_{\mathbf{r}}$\ is the total number
operator. It is easy to check that%
\begin{eqnarray}
&&h_{\mathbf{r}}^{\alpha }[e^{-i\mathbf{q\cdot r}}\hat{a}_{\mathbf{r}%
}^{\dagger }+e^{-i\mathbf{q\cdot }(\mathbf{r}+\mathbf{e}_{\alpha })}\hat{a}_{%
\mathbf{r}+\mathbf{e}_{\alpha }}^{\dagger }]\left\vert 0\right\rangle _{%
\mathbf{r}}\left\vert 0\right\rangle _{\mathbf{r}+\mathbf{e}_{\alpha }}=0, 
\nonumber \\
&&h_{\mathbf{r}}^{\alpha }\hat{a}_{\mathbf{r}}^{\dagger }\hat{a}_{\mathbf{r}+%
\mathbf{e}_{\alpha }}^{\dagger }\left\vert 0\right\rangle _{\mathbf{r}%
}\left\vert 0\right\rangle _{\mathbf{r}+\mathbf{e}_{\alpha }}=0,  \nonumber
\\
&&h_{\mathbf{r}}^{\alpha }\left\vert 0\right\rangle _{\mathbf{r}}\left\vert
0\right\rangle _{\mathbf{r}+\mathbf{e}_{\alpha }}=0,
\end{eqnarray}%
which ensures that%
\begin{equation}
H\left\vert \psi _{n}\right\rangle =n\sum\limits_{\alpha =1}^{3}V_{\alpha
}\left\vert \psi _{n}\right\rangle .
\end{equation}

{To understand the underlying mechanism for the existence of such
eigenstates, we consider a free boson model on an $N$-site ring with the
following Hamiltonian\begin{equation}
H_{\text{FB}}=\sum_{j=1}^{N}\left( \hat{b}_{j}^{\dag }\hat{b}_{j+1}+\text{H.c.}\right) ,
\end{equation}where $b_{j}^{\dag }$ and $b_{j}$\ are creation and annihilation operators
for a boson on site $j$, respectively. We consider a two-boson eigenstate
given by\begin{equation}
\left\vert 2,q\right\rangle =\left( \sum_{j=1}^{N}e^{iqj}\hat{b}_{j}^{\dag
}\right) ^{2}\left\vert 0\right\rangle ,
\end{equation}which satisfies the Schrodinger equation
\begin{equation}
H_{\text{FB}}\left\vert 2,q\right\rangle =4\cos q\left\vert 2,q\right\rangle
.  \label{Seq_FB2}
\end{equation}We note that the state $\left\vert 2,q\right\rangle $\ and the hardcore
boson state\begin{equation}
\left\vert \psi _{2}\right\rangle =\left( \sum_{j=1}^{N}e^{iqj}\hat{a}_{j}^{\dagger }\right) ^{2}\left\vert 0\right\rangle ,
\end{equation}have the relation\begin{equation}
\left\vert 2,q\right\rangle =\left\vert \psi _{2}\right\rangle
+\sum_{j=1}^{N}e^{i2qj}\left( \hat{b}_{j}^{\dag }\right) ^{2}\left\vert
0\right\rangle .
\end{equation}The Eq. (\ref{Seq_FB2}) can be explicitly written as\begin{eqnarray}
H_{\text{FB}}\left\vert 2,q\right\rangle &=&H_{\text{FB}}\left\vert \psi
_{2}\right\rangle +H_{\text{FB}}\sum_{j=1}^{N}e^{i2qj}\left( \hat{b}_{j}^{\dag }\right) ^{2}\left\vert 0\right\rangle   \nonumber \\
&=&H_{\text{FB}}\left\vert \psi _{2}\right\rangle +4\cos
q\sum_{j=1}^{N}e^{iqj}e^{iq(j+1)}\hat{b}_{j}^{\dag }\hat{b}_{j+1}^{\dag
}\left\vert 0\right\rangle .
\end{eqnarray}
The identity\begin{equation}
\hat{b}_{j}^{\dag }\hat{b}_{j+1}^{\dag }\left\vert 0\right\rangle =\left(
\hat{b}_{l}^{\dag }\hat{b}_{l}\hat{b}_{l+1}^{\dag }\hat{b}_{l+1}\right) \hat{b}_{j}^{\dag }\hat{b}_{j+1}^{\dag }\left\vert 0\right\rangle ,
\end{equation}ensures\begin{equation}
H_{\text{FB}}\left\vert 2,q\right\rangle =H_{\text{FB}}\left\vert \psi
_{2}\right\rangle +2\cos q\left( \sum_{l}^{N}\hat{b}_{l}^{\dag }\hat{b}_{l}\hat{b}_{l+1}^{\dag }\hat{b}_{l+1}\right) \left\vert 2,q\right\rangle .
\end{equation}Then we have\begin{equation}
H_{\text{FB}}\left\vert \psi _{2}\right\rangle +2\cos q\left( \sum_{l}^{N}\hat{b}_{l}^{\dag }\hat{b}_{l}\hat{b}_{l+1}^{\dag }\hat{b}_{l+1}\right)
\left\vert 2,q\right\rangle =4\cos q\left\vert 2,q\right\rangle ,
\end{equation}and furthermore\begin{equation}
\left[ H_{\text{FB}}+2\cos q\left( \sum_{l}^{N}\hat{b}_{l}^{\dag }\hat{b}_{l}\hat{b}_{l+1}^{\dag }\hat{b}_{l+1}\right) \right] \left\vert \psi
_{2}\right\rangle =4\cos q\left\vert 2,q\right\rangle .
\end{equation}Applying the projection operator $P$, given by\begin{equation}
P\left( \hat{b}_{j}^{\dag }\right) ^{2}\left\vert 0\right\rangle =0,
\end{equation}to the above equation, we obtain the corresponding Schrodinger equation for
the state $\left\vert \psi _{2}\right\rangle $. This operator can be
realized by adding infinite on-site interaction, which rules out the doublon
state $\left( \hat{b}_{j}^{\dag }\right) ^{2}\left\vert 0\right\rangle $. It
is clear that the hardcore constraint forbbids transitions from the state $\left( \hat{b}_{j}^{\dag }\right) ^{2}\left\vert 0\right\rangle $\ to states
$\hat{b}_{j-1}^{\dag }\hat{b}_{j}^{\dag }\left\vert 0\right\rangle $\ and $\hat{b}_{j}^{\dag }\hat{b}_{j+1}^{\dag }\left\vert 0\right\rangle $. To
compensate for these states, the resonant NN interaction can be introduced.
In this sense, the resonant NN interaction serves to cancel out the effects
from hardcore scattering.}

In addition, state $\left\vert \psi _{n}\right\rangle $\ possesses ODLRO due
to the fact that the correlation function 
\begin{equation}
\left\langle \psi _{n}\right\vert \hat{a}_{\mathbf{r}}^{\dagger }\hat{a}_{%
\mathbf{r}+\mathbf{R}}\left\vert \psi _{n}\right\rangle =e^{-iq\cdot\mathbf{
R}}\frac{\left( N-n\right) n}{N(N-1)},
\end{equation}%
does not decay as $\left\vert \mathbf{R}\right\vert $\ increases. The detail
derivation is given in the Appendix A. 
{In the dilute case where $n\ll N$, the on-site and NN interactions can be neglected. The
system can be regarded as a free boson system, exhibiting a superfuid state.
However, in the case where $n\sim N$, the on-site and NN
interactions are not negligible. In the extreme case, we have \begin{equation}
\left\vert \psi _{N}\right\rangle \propto \frac{1}{\Omega _{n}}\prod_{\mathbf{r}}\hat{a}_{\mathbf{r}}^{\dagger }\left\vert 0\right\rangle ,
\end{equation}which is obviously an insulating state. This can be seen from the
fact that the corresponding correlation function $\left\langle \psi
_{N}\right\vert \hat{a}_{\mathbf{r}}^{\dagger }\hat{a}_{\mathbf{r}+\mathbf{R}}\left\vert \psi _{N}\right\rangle $\ vanishes.}

\section{Coalescing condensate states}

\label{Coalescing condensate states}

{In this section, we turn to investigate the unique feature, the\ EP, in the present non-Hermitian system described by Eq. (\ref{H}). On the one
hand, the strength of non-Hermitian impurities in the resonant Hamiltonian
depends on the value of $q$. On the other hand, the discrete values
of $q$\ can be adjusted by the size of the system. It is presumbly
the case that there exist some critical values of $q$, at which the
Hamiltonian reaches the EP.}

In parallel, without loss of generality, we have 
\begin{equation}
\left\vert \varphi _{n}\right\rangle =\frac{1}{\Omega _{n}}\left( \sum_{ 
\mathbf{r}}\hat{a}_{\mathbf{r}}^{\dagger }e^{i\mathbf{q\cdot r}}\right)
^{n}\left\vert 0\right\rangle ,
\end{equation}%
for the equation 
\begin{equation}
H^{\dag }\left\vert \varphi _{n}\right\rangle =n\sum\limits_{\alpha
=1}^{3}V_{\alpha }\left\vert \varphi _{m}\right\rangle ,
\end{equation}%
which establishes the biorthonormal set $\left\{ |\varphi _{m}\rangle ,|\psi
_{n}\rangle \right\} $, satisfying 
\begin{equation}
\langle \varphi _{m}\left\vert \psi _{n}\right\rangle =\delta _{mn},
\end{equation}%
except for some special cases. We start the demonstrations\ from the
simplest case with $n=1$. Straightforward derivation shows that 
\begin{equation}
\langle \varphi _{1}\left\vert \psi _{1}\right\rangle =0,
\end{equation}%
if $q_{\alpha }=q_{\alpha }^{\mathrm{c}}=\pi m_{\alpha }/N_{\alpha }$ $%
(m_{\alpha }\in \lbrack 1,2N_{\alpha }-1],$ $m_{\alpha }\neq N_{\alpha })$\
for any\ one of $\alpha $, which indicates that the complete set of
eigenstates is spoiled. According to non-Hermitian quantum mechanics, the
Hamiltonian with parameter $q_{\alpha }^{\mathrm{c}}$\ has an EP, and $%
\left\vert \psi _{1}\right\rangle $\ is referred to as a coalescing state.
We note that an EP can be induced by the parameter along a single direction
(any\ one of $\alpha =1$, $2$, and $3$). In this sense, the conditions for
occurrence of EP are independent of three directions. Then one can
investigate the EP problem from a 1D system, which makes things easily
accessible. However, it is not a straightforward conclusion that $\left\vert
\psi _{n}\right\rangle $\ is a coalescing state simultaneously, since
operator $\hat{a}_{\mathbf{r}}^{\dagger }$\ obeys an unusual commutation
relations in Eq. (\ref{resonant condition}).

Considering a 1D system with a set of Hamiltonians in Eq. (\ref{H}) with
open boundary condition, i.e., $N_{1}=N$, $N_{2}=N_{3}=1$, and $q_{1}^{%
\mathrm{c}}=\pi m_{1}/N$ $=2\pi m_{1}/\left( 2N\right) \ (m_{1}\in \lbrack
1,2N-1],$ $m_{1}\neq N)$, each Hamiltonian $H(q_{1}^{\mathrm{c}})$\ is tuned
at EP. The matrix representation of $H(q_{1}^{\mathrm{c}})$\ in the
single-particle invariant subspace should have an EP \cite%
{jin2009solutions,zhang2013self} due to the existence of $2\times 2$\ Jordan
block. A natural question is what happens in the $n$-particle invariant
subspace and whether $\left\vert \psi _{n}\right\rangle $\ is also a
coalescing state. To answer this question, we consider another set of
Hamiltonians in Eq. (\ref{H}) with periodic boundary condition, i.e., $%
N_{1}=2N$, $N_{2}=N_{3}=1$, and $q_{1}=2\pi m_{1}/\left( 2N\right) \
(m_{1}\in \lbrack 1,2N])$. Each Hamiltonian $H(q_{1})$\ is Hermitian and
supports the Schrodinger equation%
\begin{equation}
H(q_{1})\left\vert \Psi _{n}\right\rangle =nV_{1}\left\vert \Psi
_{n}\right\rangle ,
\end{equation}%
with eigenstates%
\begin{equation}
\left\vert \Psi _{n}\right\rangle =\frac{1}{\Omega _{n}}\left(
\sum_{m=1}^{2N}\hat{a}_{m\mathbf{e}_{1}}^{\dagger }e^{-imq_{1}}\right)
^{n}\left\vert 0\right\rangle .
\end{equation}%
Notably, we find that the coalescing state $\left\vert \psi
_{1}\right\rangle $\ of $H(q_{1}^{\mathrm{c}})$\ is exactly the half part of
the eigenstate $\left\vert \Psi _{1}\right\rangle $. It is a starting point,
based on which we can show that%
\begin{equation}
\langle \varphi _{n}\left\vert \psi _{n}\right\rangle =0,  \label{zero norm}
\end{equation}%
i.e., state $\left\vert \psi _{n}\right\rangle $\ is also a coalescing state
of the Hamiltonian $H(q_{1}^{\mathrm{c}})$. The detail derivation is given
in the Appendix B. 
{It has been shown that non-Hermitian impurities
have an intimate relationship with the scattering solutions of Hermitian
systems\cite{jin2010physics,jin2011partitioning,jin2011physical}. In comparison with previous studies, the coalescing states
obtained in this work not only provide an exact example but also extend the
investigation to many-body systems.}

\begin{figure*}[tbh]
\centering
\includegraphics[width=1\textwidth]{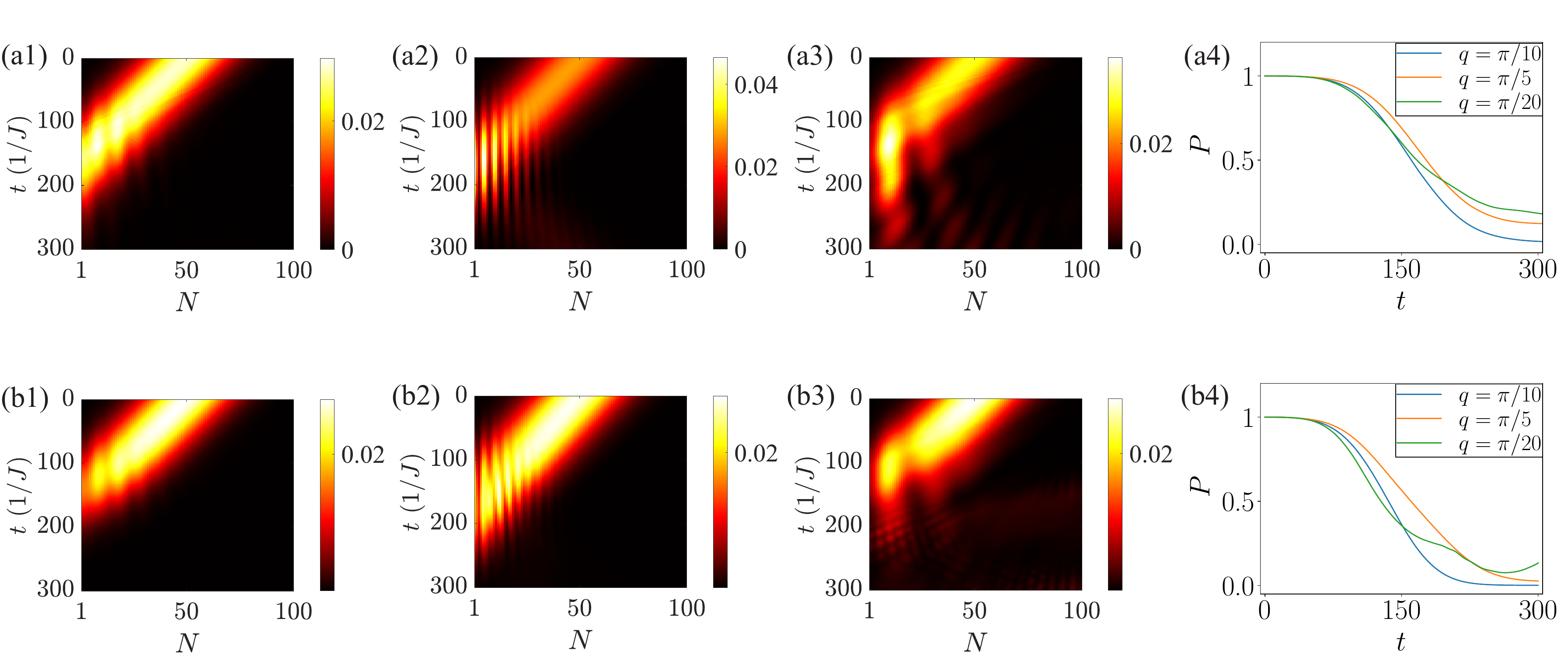}
\caption{Plots of $p_{j}(t)$ and $P(t)$ defined in Eqs. (\protect\ref{pj(t)}) and (\protect\ref{P(t)}) for the initial states defined in Eq. (\protect
\ref{initial GW2}) with $n=1$ in (a1)-(a3) and $n=2$ in (b1)-(b3),
respectively. The parameters of the Gaussian wavepacket are $\protect\alpha %
=0.05 $, $N_{0}=50$ and $q=\protect\pi /10$ in (a1, b1), $\protect\pi /5$ in
(a2, b2), $\protect\pi /20$\ in (a3, b3). The system parameter is $q_{\text{%
c }}=\protect\pi /10$. The profiles of $p_{j}(t)$ in (a2, b2, a3, b3)\
exhibit evident interference fringes indicating reflections from the end of
the chain.\ The plots of $P(t)$\ indicate the perfect probability absorption
for the resonant incident wavepackets.}
\label{fig1}
\end{figure*}

\section{Resonant scattering of non-Hermitian impurity}

\label{Resonant scattering of non-Hermitian impurity}

In this section, we focus on our study on 1D system for simplicity. The
obtained result can be extended to 2D and 3D systems. We start with our
investigation on the Hamiltonian in a single-particle invariant subspace, in
which the single-particle dynamics obeys a free boson model with the $%
\mathcal{PT}$ symmetric\ non-Hermitian Hamiltonian 
\begin{equation}
H_{\text{FB}}=\sum_{j=1}^{N-1}\left( \hat{b}_{j}^{\dag }\hat{b}_{j+1}+\text{%
H.c.}\right) +e^{-iq}\hat{b}_{1}^{\dag }\hat{b}_{1}+e^{iq}\hat{b}_{N}^{\dag }%
\hat{b}_{N},  \label{H_1}
\end{equation}%
on an $N$-site chain with complex on-site potential at two ends. We take a
dimensionless constant\ for the sake of simplicity. According to the
analysis in last two sections, states%
\begin{equation}
\left( \sum_{j=1}^{N}e^{iqj}\hat{b}_{j}^{\dag }\right) ^{n}\left\vert
0\right\rangle ,
\end{equation}%
are $n$-boson eigenstates of $H_{\text{FB}}$. In addition, these states with
different $n$ are coalescing states under the condition $q=q_{\text{c}}=\pi
m/N\ (m\in \lbrack 1,2N-1],$ $m\neq N)$. Furthermore, this constraint for $%
q_{\text{c}}$ is satisfied automatically in large $N$ limit. It has been
shown that this fact has intimate connection to the reflectionless
scattering problem \cite{zhang2013self} for the semi-infinite system%
\begin{equation}
H_{\text{FB}}^{\infty }=\sum_{j=1}^{\infty }\left( \hat{b}_{j}^{\dagger }%
\hat{b}_{j+1}+\text{H.c.}\right) +e^{-iq_{\text{c}}}\hat{b}_{1}^{\dagger }%
\hat{b}_{1},
\end{equation}%
with a complex impurity at the end. The dynamic demonstration of this exact
result is the near-perfect reflectionless of a Gaussian wavepacket with
resonant momentum $q=q_{\text{c}}$. Specifically, we consider an initial $n$%
-boson state in the form%
\begin{equation}
\left\vert \phi (0)\right\rangle =\left( \sum_{j}g_{j}\hat{b}_{j}^{\dag
}\right) ^{n}\left\vert 0\right\rangle ,  \label{initial GW}
\end{equation}%
where the single-boson wave function has the form%
\begin{equation}
g_{j}=e^{-\frac{^{\alpha ^{2}}}{2}(j-N_{0})^{2}}e^{iqj}.
\end{equation}%
The shape and center of the wavepacket are determined by parameters $\alpha $%
\ and $N_{0}$.\ The near-perfect reflectionless indicates the time evolution
of $\left\vert \phi (0)\right\rangle $ obeys\ 
\begin{equation}
\lim_{t\rightarrow \infty }e^{-iH_{\text{FB}}^{\infty }t}\left\vert \phi
(0)\right\rangle \approx 0.
\end{equation}%
It holds true for any $n$ with small $\alpha $, due to the following facts.
(i) A wider single-boson wavepacket with $q=q_{\text{c}}$ can reflect the
exact result for plane wave scattering from the end \cite{zhang2013self}.
(ii) Multi-boson wavepacket shares the same dynamic behavior of a
single-boson, since there is no interaction between bosons.

Now we turn to the hardcore boson Hubbard model, by ruling out the double
occupation\ and\ adding the resonant NN interaction with the Hamiltonian%
\begin{equation}
H_{\text{HB}}^{\infty }=\sum_{j=1}^{\infty }\left( \hat{a}_{j}^{\dagger }%
\hat{a}_{j+1}+\text{H.c.}+2\cos q_{\text{c}}\hat{n}_{j}\hat{n}_{j+1}\right)
+e^{-iq_{\text{c}}}\hat{a}_{1}^{\dagger }\hat{a}_{1}.
\end{equation}%
The question is whether we still have the result 
\begin{equation}
\lim_{t\rightarrow \infty }e^{-iH_{\text{HB}}^{\infty }t}\left( \sum_{j}g_{j}%
\hat{a}_{j}^{\dagger }\right) ^{n}\left\vert 0\right\rangle \approx 0,
\end{equation}%
for the initial state%
\begin{equation}
\left\vert \phi (0)\right\rangle =\left( \sum_{j}g_{j}\hat{a}_{j}^{\dagger
}\right) ^{n}\left\vert 0\right\rangle .  \label{initial GW2}
\end{equation}%
To answer this question, numerical simulations are performed for the $n$%
-hardcore-boson initial wavepackets $\left\vert \phi (0)\right\rangle $\
with $q$ around $q_{\text{c}}$. 
{The evolved states $\left\vert \phi
\left( t\right) \right\rangle =e^{-iH_{\text{HB}}t}\left\vert \phi
(0)\right\rangle $\ are computed by exact diagonalization for finite
systems\ with several typical set of parameters.}

The profiles of the evolved states and the total probabilities are measured
by%
\begin{equation}
p_{j}(t)=\left\vert \frac{\hat{a}_{j}\left\vert \phi (t)\right\rangle }{%
\left\vert \phi (0)\right\rangle }\right\vert ^{2}=\left\vert \frac{\hat{a}%
_{j}e^{-iH_{\text{HB}}^{\infty }t}\left\vert \phi (0)\right\rangle }{%
\left\vert \phi (0)\right\rangle }\right\vert ^{2},  \label{pj(t)}
\end{equation}%
and%
\begin{equation}
P(t)=\frac{1}{n}\sum_{j}p_{j}(t)=\left\vert \frac{e^{-iH_{\text{HB}}^{\infty
}t}\left\vert \phi (0)\right\rangle }{\left\vert \phi (0)\right\rangle }%
\right\vert ^{2},  \label{P(t)}
\end{equation}%
respectively.

We plot the two quantities $p_{j}(t)$\ and\ $P(t)$ in Fig. \ref{fig1}(a) and
(b) for selected systems with two types of initial states. The parameters of
the initial wavepackets and the driven systems are given in the captions of
the figure. We consider the driven Hamiltonians\ on a finite site chain with
the parameter $q_{\text{c}}=\pi /10$. The initial states are taken as: (a) a
single-boson wavepacket and (b) a two-boson wavepacket, respectively. The
center momenta of the wavepackets are taken as $q=q_{\text{c}}$, $q_{\text{c}%
}/2$, and $2q_{\text{c}}$,\ respectively. According to the analysis above,
both the single- and two-boson incident wavepackets with the resonant
momentum $q_{\text{c}}$\ should be approximately absorbed by the ending
non-Hermitian impurity. From Fig. \ref{fig1}(a1)-(a3), one can observe that
there is no reflection pattern in (a1) for the resonant wavepacket. In
contrast, evident interference fringes are observed near the end of the
chain in cases (a2) and (a3), where the incident wavepackets are
off-resonant. Such patterns result from the interference between incident
and reflected wavepackets. From Fig. \ref{fig1}(a4), one can see that the
probabilities for each case reduce significantly when the wavepackets reach
the impurity, exhibiting absorption behavior. Notably, the absorption is
minimal when the initial wavepacket carries the resonant momentum $q_{\text{c%
}}$. This result is consistent with the patterns observed in Fig. \ref{fig1}(a1)-(a3).
Actually, the plots in Fig. \ref{fig1}(a1)-(a4) are provided to compare the
results with those for two-boson incident wavepackets, which are shown in %
Fig. \ref{fig1}(b1)-(b4). We observe that the two sets of patterns have the
similar profiles. The results indicate perfect absorption for resonant
incident wavepackets with $q=q_{\text{c}}$\textbf{\ }in both the one- and
two-boson cases, which is in accordance with the previous theoretical
analysis. Then, on the one hand, the results indicate that the underlying
mechanism for the coalescing condensate states in an open chain is the
reflectionless absorption of many-particle wavepackets with resonant
momentum by the non-Hermitian boundary.\textbf{\ }{These resonant
absorptions demonstrate the dynamic signature of the EP. }On the other hand,
it reveals an interesting dynamic behavior of a many-particle wavepacket: it
behaves as a single-particle wavepacket. The NN resonant interaction {indeed
plays a role in canceling out the scattering effect arising from the
hardcore interaction.}

\begin{table}[tbp]
\caption{The structures of energy levels for $10$-site open chain with
different $q$ and filling particle number $n$.\ We list the numbers of
coalescing states $n_{\text{CS}}$, of order $n_{\text{OR}}$,\ and the
numbers of complex levels $n_{\text{CM}}$ in the form ($n_{\text{CM}}$, $n_{%
\text{OR}}\times n_{\text{CS }}$). It indicates that all the energy levels
are real and all the systems contain $2$-order coalescing states.}
\label{Table I}%
\begin{tabular}{ccccccccccccccc}
\hline\hline
q &  &  & n=2 &  &  & 3 &  &  & 4 &  &  & 5 &  &  \\ \hline
$\pi/10$ &  &  & $0,2\times1$ &  &  & $0,2\times1$ &  &  & $0,2\times1$ &  & 
& $0,2\times1$ &  &  \\ 
$2\pi/10$ &  &  & $0,2\times1$ &  &  & $0,2\times1$ &  &  & $0,2\times36$ & 
&  & $0,2\times43$ &  &  \\ 
$3\pi/10$ &  &  & $0,2\times1$ &  &  & $0,2\times1$ &  &  & $0,2\times1$ & 
&  & $0,2\times1$ &  &  \\ 
$4\pi/10$ &  &  & $0,2\times1$ &  &  & $0,2\times1$ &  &  & $0,2\times36$ & 
&  & $0,2\times43$ &  &  \\ \hline\hline
\end{tabular}%
\end{table}
\begin{table}[tbp]
\caption{The same as Table \protect\ref{Table I}\ but for $N_{1}\times
N_{2}=5\times 3$ lattice with different ($q_{1}$, $q_{2}$). We take open
boundary condition in $q_{1}$-direction and periodic boundary condition in $%
q_{2}$-direction. It indicates that all the systems contain complex levels
and multiple $2$-order coalescing states.}
\label{Table II}%
\begin{tabular}{ccccccccccccccc}
\hline\hline
($q_1,q_2$) &  &  & n=2 &  &  & 3 &  &  & 4 &  &  &  &  &  \\ \hline
$(\pi/5,2\pi/3)$ &  &  & $7,2\times11$ &  &  & $135,2\times5$ &  &  & $%
525,2\times2$ &  &  &  &  &  \\ 
$(2\pi/5,2\pi/3)$ &  &  & $16,2\times11$ &  &  & $173,2\times5$ &  &  & $%
586,2\times2$ &  &  &  &  &  \\ 
$(3\pi/5,2\pi/3)$ &  &  & $16,2\times11$ &  &  & $171,2\times5$ &  &  & $%
586,2\times2$ &  &  &  &  &  \\ 
$(4\pi/5,2\pi/3)$ &  &  & $13,2\times11$ &  &  & $159,2\times5$ &  &  & $%
570,2\times2$ &  &  &  &  &  \\ \hline\hline
\end{tabular}%
\end{table}
\begin{figure*}[tbh]
\centering
\includegraphics[width=1\textwidth]{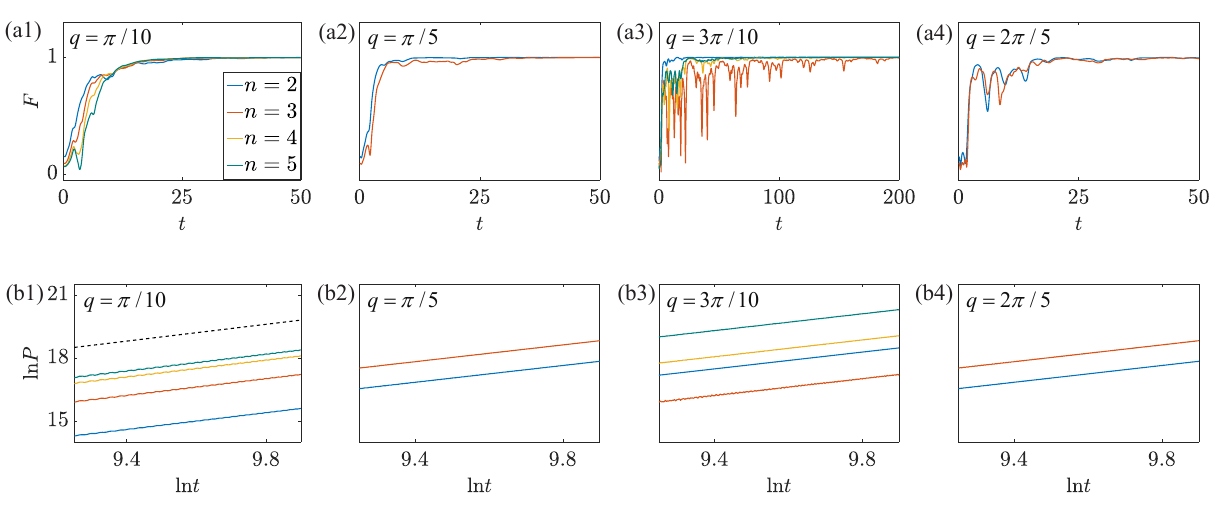}
\caption{(a1)-(a4) present the time evolution of fidelity of $n$-particle
one-dimensional system with $q=m\protect\pi /N$ where m ranges from $1$ to $%
4 $. (b1)-(b4) present the time evolution of probability of the same system.
The slope of dashed line as reference is $2$ and it is nearly parallel to
other colored lines. Here we only plot the results with a single 2 order
coalescing state. The initial state is $\prod_{j=1}^{n}\hat{a}_{j}^{\dagger
}\left\vert 0\right\rangle $ and the size of system $N=10$, where the number
n is indicated in the panel. {The results show that the unentangled
initial states can indeed evolve to the corresponding target with high
fidelity.}}
\label{fig2}
\end{figure*}

\section{Dynamic generation of condensate states}

\label{Dynamic generation of condensate states}

Condensate state as macroscopic quantum state has significance both in
theoretical and experimental physics. The stability of such states is
crucial in practice. In general, it can be prepared as ground state by
decreasing the temperature. Dynamical generation of nonequilibrium steady
condensate states is another way, which has been received much attention
recently. The advantage of an EP system is that a coalescing state is also a
target state for a long-time evolution of various initial states. In the
above, we have shown that the state $\left\vert \psi _{n}\right\rangle $\ is
also a coalescing state of the Hamiltonian $H(q_{1}^{\mathrm{c}})$. However,
the order of EP is unknown, maybe depends on the particle numbers $n$, and
the numbers of dimensions $n_{\mathrm{d}}$\ at which the resonant boundary
is taken.

In contrast, the order of EP for free boson model can be exactly obtained.
In a single-boson invariant subspace, the maximal order of EP is $n_{\mathrm{d}}$, and then is $m$ for $n$-boson system,where 
\begin{equation}
m=\frac{(n_\mathrm{d}+n-1)!}{n!(n_\mathrm{d}-1)!}.
\end{equation}
According to the non-Hermitian quantum theory, there is a $m$-D Jordan block 
$M$ in the matrix representation of the Hamiltonian, which ensures that%
\begin{equation}
M^{m}=0.
\end{equation}%
The dynamics for any state in this subspace of Jordan block, referred to as
auxiliary states,\ is governed by the time evolution operator%
\begin{equation}
U(t)=e^{-iEt}e^{-iMt}=e^{-iEt}\sum_{l=0}^{m-1} \frac{1}{l!}\left(
-iMt\right) ^{l},
\end{equation}%
where $E$ is a constant without any effect on the evolved state. It
indicates that for an initial state $\left\vert \phi (0)\right\rangle $
involving the auxiliary states we have $\left\vert \phi \left( t\right)
\right\rangle =U(t)\left\vert \phi (0)\right\rangle $ 
\begin{equation}
\lim_{t\rightarrow \infty }\left\vert \phi (t)\right\rangle \propto \left(
\sum_{\mathbf{r}}e^{-i\mathbf{q\cdot r}}\hat{b}_{\mathbf{r}}^{\dag }\right)
^{n}\left\vert 0\right\rangle ,
\end{equation}%
within large $t$\ region and 
\begin{equation}
\left\vert \left\vert \phi \left( t\right) \right\rangle \right\vert
^{2}\propto t^{2(m-1)},
\end{equation}%
which is also a dynamic demonstration for the order of the Jordan block.

However, such an analysis may be invalid due to the particle-particle
interactions, i.e., replacing $\hat{b}_{\mathbf{r}}^{\dag }$\ by $\hat{a}_{%
\mathbf{r}}^{\dag }$. In this situation, numerical simulations for finite
size systems can shed some light on the dynamic generation of the condensate
states of hardcore bosons.

{We perform numerical simulations on finite systems with the following
considerations. (i) The analysis above only predicts the results within
large time domain. The efficiency of the scheme should be estimated from
numerical simulations. (ii) The final state seems to be independent of the
initial states, }which can be simply unentangled $n$-boson initial states in
the form $\left\vert \phi (0)\right\rangle =\prod_{\left\{
l_{1},l_{2},...l_{j},...l_{n},\right\} }\hat{a}_{l_{j}}^{\dagger }\left\vert
0\right\rangle $. (iii) The order of EP can be observed by the dynamic
process. The energy levels of many-particle non-Hermitian system are
complicated, containing complex energy levels and multiple coalescing
states, which may pose an obstacle in the calculation of time evolution. Our
strategy has two steps. First, we find out the structure of energy level for
sample systems. Second, select several coalescing states as target states.
In Tables \ref{Table I} and \ref{Table II}, the numbers of complex energy
and coalescing levels are listed. 
{In general, the structure of
energy levels is characterized by three integers, the number of coalescing
states $n_{\text{CS}}$, the order $n_{\text{OR}}$,\ and
the number of complex levels $n_{\text{CM}}$. These results guide
numerical simulations. To demonstrate this scheme, the system should contain
only real levels and a coalescing state. We select several cases that
satisfy the condition to perform the computations.}

The evolved states $\left\vert \phi \left( t\right) \right\rangle $\ are
also computed by exact diagonalization for finite systems\ with several
typical set of parameters. We focus on the Dirac probability $P(t)$ defined
in Eq. (\ref{P(t)}) and the fidelity%
\begin{equation}
F(t)=\frac{\left\vert \langle \psi _{n}\left\vert \phi \left( t\right)
\right\rangle \right\vert ^{2}}{\left\vert \left\vert \phi \left( t\right)
\right\rangle \right\vert ^{2}},  \label{fidelity}
\end{equation}%
which is the measure of the distance between the final state and the
condensate states $\left\vert \psi _{n}(q)\right\rangle $. We plot the
fidelity $F(t)$ in Fig. \ref{fig2} as a function of $t$\ for selected
systems and many-particle initial states. We consider the driven
Hamiltonians on $10$-site chains with the system parameters $q=m\pi /10$,
where $m$ ranges from $1$ to $4$. For $m=1$ and $3$, we consider the sectors
with $n=2$, $3$, $4$, and $5$.\ For $m=2$ and $4$, we only consider the
sectors with $n=2$ and $3$.\ According to the results listed in Tables \ref%
{Table I}, the corresponding energy levels are all real and the systems
contain only one $2$-order EP, with the coalescing condensate states in the
form%
\begin{equation}
\left\vert \psi _{n}\right\rangle =\frac{1}{\Omega _{n}}\left(
\sum_{j=1}^{10}\hat{a}_{j}^{\dagger }e^{-im\pi j/10}\right) ^{n}\left\vert
0\right\rangle .  \label{target N=10}
\end{equation}%
The $n$-boson initial states are taken in the explicit form%
\begin{eqnarray}
\left\vert \phi \left( 0\right) \right\rangle  =\hat{a}_{1}^{\dagger
}\left\vert 0\right\rangle ,\hat{a}_{1}^{\dagger }\hat{a}_{2}^{\dagger
}\left\vert 0\right\rangle ,\hat{a}_{1}^{\dagger }\hat{a}_{2}^{\dagger }\hat{%
a}_{3}^{\dagger }\left\vert 0\right\rangle , \hat{a}_{1}^{\dagger }\hat{a}_{2}^{\dagger }\hat{a}_{3}^{\dagger }\hat{a}%
_{4}^{\dagger }\left\vert 0\right\rangle ,\hat{a}_{1}^{\dagger }\hat{a}%
_{2}^{\dagger }\hat{a}_{3}^{\dagger }\hat{a}_{4}^{\dagger }\hat{a}%
_{5}^{\dagger }\left\vert 0\right\rangle ,
\end{eqnarray}%
which is a trivial unentangled state and can be easily prepared in the
experiment. From Fig. \ref{fig2}(a1)-(a4), one can see that the fidelities $%
F(t)$ obtained by numerical simulations for these selected cases approach $1$
as time increases. The fact that $F(t)=1$ indicates that the evolved state
is identical to the target state. In addition, we also plot the probability $%
P(t)$\ as a function of time $t$ to demonstrate the EP dynamic behavior.
According to the analysis above, the probability should obey the relation $%
P(t)\sim t^{2}$.\ From Fig. \ref{fig2} (b1)-(b4), one can see that the plots
are close to the lines in the ln$P(t)$-ln$t$\ plane, and the slopes are all
approximately equal to $2$. On the one hand, the results indicate that the
target states given in Eq. (\ref{target N=10}) can be obtained throught the
time evolution of the given initial state $\left\vert \phi \left( 0\right)
\right\rangle $. On the other hand, the underlying mechanism of such dynamic
preparation of a coalescing condensate state is the $2$-order EP dynamics,
which exhibit parabolas in the probability growth.

\section{Summary}

\label{Summary}

In summary, we have studied the Hermitian and non-Hermitian extended
hardcore Bose-Hubbard model on one-, two-, and three-dimensional\ lattices.
A set of exact eigenstates are constructed and have the following
implications: (i) The strong on-site repulsion and nearest neighboring
interaction cannot block the formation of BEC under the moderate particle
density, when two interacting strengths are matched with each other. (ii)
The solutions for Hermitian systems with periodic boundary condition are
available for any given size, in which the momentum of the condensate is
nothing but the reciprocal vector. Then the resonant non-Hermitian
impurities can result in coalescing hardcore-boson condensate states. As an
alternative stable state beyond the ground state, a coalescing state may be
obtained via natural time evolution, although it is also the excited
eigenstate of the system. In this sense, our finding not only reveals the
possible condensation of interaction bosons, but also provides a method for
condensate state engineering in an alternative way.

\section*{Acknowledgements}

% TODO: include author contributions

% TODO: include funding information

\paragraph{Funding information}

This work was supported by the National Natural Science Foundation of China
(under Grant No. 12374461).

\begin{appendix}
\numberwithin{equation}{section}

\section*{A\quad Condensate eigenstates with ODLRO}
\addcontentsline{toc}{section}{A\quad Condensate eigenstates with ODLRO}
In this appendix, we present the derivations on the eigenstates of the
Hamiltonian in Eq. (\ref{H}) and resonant conditions for many-body
coalescing states.
\label{Appendix1}
Consider the state

\begin{equation}
	\tag{A.1}
	\left\vert \psi _{n}\right\rangle =\frac{1}{\Omega _{n}}\left( [e^{-i\mathbf{%
			q\cdot r}}\hat{a}_{\mathbf{r}}^{\dagger }+e^{-i\mathbf{q\cdot }(\mathbf{r}+%
		\mathbf{R})}\hat{a}_{\mathbf{r}+\mathbf{R}}^{\dagger }]+A\right)
	^{n}\left\vert 0\right\rangle ,
\end{equation}%
where $A$\ is an operator which does not contain $\hat{a}_{\mathbf{r}}^{\dagger }$%
\ and $\hat{a}_{\mathbf{r}+\mathbf{R}}^{\dagger }$. We have%
\begin{equation}
	\tag{A.2}
	\left\vert \psi _{n}\right\rangle =\frac{1}{\Omega _{n}}%
	\sum_{k=1}^{n}C_{n}^{k}A^{n-k}[e^{-i\mathbf{q\cdot r}}\hat{a}_{\mathbf{r}%
	}^{\dagger }+e^{-i\mathbf{q\cdot }(\mathbf{r}+\mathbf{R})}\hat{a}_{\mathbf{r}%
		+\mathbf{R}}^{\dagger }]^{n}\left\vert 0\right\rangle .
\end{equation}%
In fact, we note that%
\begin{equation}
	\tag{A.3}
	\lbrack e^{-i\mathbf{q\cdot r}}\hat{a}_{\mathbf{r}}^{\dagger }+e^{-i\mathbf{%
			q\cdot }(\mathbf{r}+\mathbf{R})}\hat{a}_{\mathbf{r}+\mathbf{R}}^{\dagger
	}]^{2}\left\vert 0\right\rangle =2e^{-i\mathbf{q\cdot }(2\mathbf{r}+\mathbf{R%
		})}\hat{a}_{\mathbf{r}}^{\dagger }\hat{a}_{\mathbf{r}+\mathbf{R}}^{\dagger
	}\left\vert 0\right\rangle ,  \notag
\end{equation}%
but%
\begin{equation}
	\tag{A.4}
	\lbrack e^{-i\mathbf{q\cdot r}}\hat{a}_{\mathbf{r}}^{\dagger }+e^{-i\mathbf{%
			q\cdot }(\mathbf{r}+\mathbf{R})}\hat{a}_{\mathbf{r}+\mathbf{R}}^{\dagger
	}]^{k}\left\vert 0\right\rangle =0,(k>2).
\end{equation}%
Then%
\begin{align}
	\left\vert \psi _{n}\right\rangle =&\frac{1}{\Omega _{n}}%
	\{A^{n}+nA^{n-1}[e^{-i\mathbf{q\cdot r}}\hat{a}_{\mathbf{r}}^{\dagger }+e^{-i%
		\mathbf{q\cdot }(\mathbf{r}+\mathbf{R})}\hat{a}_{\mathbf{r}+\mathbf{R}%
	}^{\dagger }]  \notag \\
	&+e^{-i\mathbf{q\cdot }(2\mathbf{r}+\mathbf{R})}n(n-1)A^{n-2}\hat{a}_{%
		\mathbf{r}}^{\dagger }\hat{a}_{\mathbf{r}+\mathbf{R}}^{\dagger }\}\left\vert
	0\right\rangle\tag{A.5} .
\end{align}%

One can take $\mathbf{R=e}_{\alpha }$, and then we have%
\begin{equation}
	\tag{A.6}
	h_{\mathbf{r}}^{\alpha }\left\vert \psi _{n}\right\rangle =0,
\end{equation}%
which ensures that%
\begin{equation}
	\tag{A.7}
	H\left\vert \psi _{n}\right\rangle =n\sum\limits_{\alpha =1}^{3}V_{\alpha
	}\left\vert \psi _{n}\right\rangle .
\end{equation}%
Furthermore, we have%
\begin{align}
	\hat{a}_{\mathbf{r}}^{\dagger }\hat{a}_{\mathbf{r}+\mathbf{R}}\left\vert
	\psi _{n}\right\rangle =&\frac{1}{\Omega _{n}}\left\{nA^{n-1}[e^{-i\mathbf{%
			q\cdot r}}\hat{a}_{\mathbf{r}}^{\dagger }\hat{a}_{\mathbf{r}+\mathbf{R}}\hat{%
		a}_{\mathbf{r}}^{\dagger }  +e^{-i\mathbf{q\cdot }(\mathbf{r}+\mathbf{R})}\hat{a}_{\mathbf{r}%
	}^{\dagger }\hat{a}_{\mathbf{r}+\mathbf{R}}\hat{a}_{\mathbf{r}+\mathbf{R}%
	}^{\dagger }]\right\}\left\vert 0\right\rangle  \notag \\
	=&\frac{1}{\Omega _{n}}nA^{n-1}e^{-i\mathbf{q\cdot }(\mathbf{r}+\mathbf{R})}%
	\hat{a}_{\mathbf{r}}^{\dagger }\left\vert 0\right\rangle\tag{A.8} ,
\end{align}%
which results in the correlation function%
\begin{align}
	\left\langle \psi _{n}\right\vert \hat{a}_{\mathbf{r}}^{\dagger }\hat{a}_{%
		\mathbf{r}+\mathbf{R}}\left\vert \psi _{n}\right\rangle =&\left( \frac{n}{%
		\Omega _{n}}\right) ^{2}e^{-i\mathbf{q\cdot R}}\left\vert A^{n-1}\left\vert
	0\right\rangle \right\vert ^{2}  \notag \\
	=&e^{-i\mathbf{q\cdot R}}\frac{\left( N-n\right) n}{N(N-1)}.\tag{A.9}
\end{align}%
We find that%
\begin{equation}
	\tag{A.10}
	\lim_{\left\vert \mathbf{R}\right\vert \mathbf{\rightarrow \infty }%
	}\left\vert \left\langle \psi _{n}\right\vert \hat{a}_{\mathbf{r}}^{\dagger }%
	\hat{a}_{\mathbf{r}+\mathbf{R}}\left\vert \psi _{n}\right\rangle \right\vert
	=\frac{n\left( N_{1}N_{2}N_{3}-n\right) }{N_{1}N_{2}N_{3}\left(
		N_{1}N_{2}N_{3}-1\right) },
\end{equation}%
which is finite number, indicating off-diagonal long-range order (ODLRO)
according to \cite{yang1962concept}.

\section*{B\quad Coalescing condensate states}
\addcontentsline{toc}{section}{B\quad Coalescing condensate states}
\label{Appendix2}
We start with the case with $n=1$. The biorthogonal norm for state $%
\left\vert \psi _{1}\right\rangle $\ is%
\begin{align}
	\langle \varphi _{1}\left\vert \psi _{1}\right\rangle =&\frac{1}{\Omega
		_{n}^{2}}\sum_{\mathbf{r}}e^{-i2\mathbf{q\cdot r}}  \notag \\
	=&\frac{1}{\Omega _{n}^{2}}\prod_{\alpha =1,2,3}\sum_{m_{\alpha
	}}e^{-i2q_{\alpha }m_{\alpha }}\tag{B.1}.
\end{align}%
We note that if
\begin{equation}
	\tag{B.2}
	\sum_{m_{\alpha }}e^{-i2q_{\alpha }m_{\alpha }}=0,
\end{equation}%
for one of $\alpha $, we have
\begin{equation}
	\tag{B.3}
	\langle \varphi _{1}\left\vert \psi _{1}\right\rangle =0,
\end{equation}%
i.e., an EP can be induced by the parameter along a single direction.

In the following, we will show that it is also true for the case with $n>1$.
We only consider 1D systems for the sake of simplicity. We focus on two
Hamiltonians. The first one is%
\begin{align}
	H_{\text{HB}}^{1} =&\sum_{j=1}^{N-1}\left( \hat{a}_{j}^{\dagger }\hat{a}%
	_{j+1}+\text{H.c.}+\cos q\hat{n}_{j}\hat{n}_{j+1}\right) +e^{-iq}\hat{a}_{1}^{\dagger }\hat{a}_{1}+e^{iq}\hat{a}_{N}^{\dagger }\hat{%
		a}_{N}\tag{B.4},
\end{align}%
which is non-Hermitian and reduced from Eq. (\ref{H}) for an $N$-site chain.
The second one is
\begin{equation}
	\tag{B.5}
	H_{\text{HB}}^{2}=\sum_{j=1}^{2N}\left( \hat{a}_{j}^{\dagger }\hat{a}_{j+1}+%
	\text{H.c.}+\cos q\hat{n}_{j}\hat{n}_{j+1}\right) ,
\end{equation}%
which is Hermitian and reduced from Eq. (\ref{H}) for a $2N$-site ring.
Defining a set of collective operators

\begin{equation}
	\tag{B.6}
	A^{+}=\frac{1}{\sqrt{N}}\sum_{j=1}^{N}\hat{a}_{j}^{\dagger }e^{-iqj},B^{+}=%
	\frac{1}{\sqrt{N}}\sum_{j=N+1}^{2N}\hat{a}_{j}^{\dagger }e^{-iqj},
\end{equation}%
and%
\begin{equation}
	\tag{B.7}
	A^{-}=\frac{1}{\sqrt{N}}\sum_{j=1}^{N}\hat{a}_{j}e^{-iqj},B^{-}=\frac{1}{%
		\sqrt{N}}\sum_{j=N+1}^{2N}\hat{a}_{j}e^{-iqj},
\end{equation}%
with $q=2\pi m/\left( 2N\right) \ (m\in \lbrack 1,2N-1],$ $m\neq N)$, a
subset of the eigenstates of $H_{\text{HB}}^{1}$\ can be expressed as%
\begin{equation}
	\tag{B.8}
	\left\vert \psi _{n}\right\rangle =\frac{1}{\Lambda _{n}}\left( A^{+}\right)
	^{n}\left\vert 0\right\rangle ,
\end{equation}%
while states%
\begin{equation}
	\tag{B.9}
	\left\vert \Psi _{n}\right\rangle =\frac{1}{{\Lambda} _{n}^{\prime}}\left(
	A^{+}+B^{+}\right) ^{n}\left\vert 0\right\rangle ,
\end{equation}%
and
\begin{equation}
	\tag{B.10}
	\left\vert \Psi _{n}^{\ast }\right\rangle =\frac{1}{{\Omega} _{n}^{\prime}}%
	\left[ \left( A^{+}+B^{+}\right) ^{\ast }\right] ^{n}\left\vert
	0\right\rangle ,
\end{equation}%
are a subset of the eigenstates of $H_{\text{HB}}^{2}$, here
\begin{equation}
	\tag{B.11}
	\Lambda_n=\frac{N^{n/2}}{(n!)\sqrt{C_N^n}}, \Lambda^{\prime}_n=\frac{N^{n/2}}{%
		(n!)\sqrt{C_{2N}^n}}.
\end{equation}
We note that state $\left\vert \psi _{n}\right\rangle $\ is a part of state $%
\left\vert \Psi _{n}\right\rangle $, which is crucial for the following
proof. The orthogonality of two states $\left\vert \Psi _{n}\right\rangle $\
and $\left\vert \Psi _{n}^{\ast }\right\rangle $\ leads to%
\begin{align}
	&\langle \Psi _{n}^{\ast }\left\vert \Psi _{n}\right\rangle  \notag \\
	=&\sum_{m=0}^{n}p_{m}\left\langle 0\right\vert \left( A^{-}\right)
	^{n-m}\left( B^{-}\right) ^{m}\left( A^{+}\right) ^{n-m}\left( B^{+}\right)
	^{m}\left\vert 0\right\rangle  \notag \\
	=&0\tag{B.12}.
\end{align}%
Taking $n=1$, we have%
\begin{equation}
	\tag{B.13}
	\left\langle 0\right\vert \left( A^{-}A^{+}+B^{-}B^{+}\right) \left\vert
	0\right\rangle =0,
\end{equation}%
which resuts in
\begin{equation}
	\tag{B.14}
	\left\langle 0\right\vert A^{-}A^{+}\left\vert 0\right\rangle =\left\langle
	0\right\vert B^{-}B^{+}\left\vert 0\right\rangle =0,
\end{equation}%
due to the translational symmetry of state $\left\vert \Psi
_{n}\right\rangle $. Based on this conclusion, taking $n=2$, we have $%
\left\langle 0\right\vert \left( B^{-}B^{+}\right) ^{2}\left\vert
0\right\rangle $ $=\left\langle 0\right\vert \left( A^{-}A^{+}\right)
^{2}\left\vert 0\right\rangle =0$. Furthermore, it turns out that
\begin{equation}
	\tag{B.15}
	\left\langle 0\right\vert \left( A^{-}A^{+}\right) ^{m}\left\vert
	0\right\rangle =0,
\end{equation}%
for $m\in \lbrack 0,n]$, which results in%
\begin{equation}
	\tag{B.16}
	\langle \varphi _{n}\left\vert \psi _{n}\right\rangle =0.
\end{equation}
\end{appendix}


% Provide your bibliography here. You have two options:

%%% FIRST OPTION
% Write your entries here directly, following the example below, including:
% Author(s), Title, Journal Ref. with year in parentheses at the end, followed by the DOI number.

%\begin{thebibliography}{99}
%\bibitem{1931_Bethe_ZP_71} H. A. Bethe, {\it Zur Theorie der Metalle. i. Eigenwerte und Eigenfunktionen der linearen Atomkette}, Zeit. f{\"u}r Phys. {\bf 71}, 205 (1931), \doi{10.1007\%2FBF01341708}.
%\bibitem{arXiv:1108.2700} P. Ginsparg, {\it It was twenty years ago today... }, \url{http://arxiv.org/abs/1108.2700}.
%\end{thebibliography}

% Use your bibtex library, formatted by the SciPost style file.
\bibliographystyle{plain}
\bibliography{SciPost_Example_BiBTeX_File.bib}

\begin{thebibliography}{10}
\providecommand{\url}[1]{\texttt{#1}}
\providecommand{\urlprefix}{URL }
\expandafter\ifx\csname urlstyle\endcsname\relax
  \providecommand{\doi}[1]{doi:\discretionary{}{}{}#1}\else
  \providecommand{\doi}{doi:\discretionary{}{}{}\begingroup
  \urlstyle{rm}\Url}\fi
\providecommand{\eprint}[2][]{\url{#2}}

\bibitem{bloch2012quantum}
I.~Bloch, J.~Dalibard and S.~Nascimbene,
\newblock \emph{{Quantum simulations with ultracold quantum gases}},
\newblock Nature Physics \textbf{8}(4), 267 (2012),
\newblock \doi{/10.1038/nphys2259}.

\bibitem{atala2014observation}
M.~Atala, M.~Aidelsburger, M.~Lohse, J.~T. Barreiro, B.~Paredes and I.~Bloch,
\newblock \emph{Observation of chiral currents with ultracold atoms in bosonic
  ladders},
\newblock Nature Physics \textbf{10}(8), 588 (2014),
\newblock \doi{/10.1038/nphys2998}.

\bibitem{aidelsburger2015measuring}
M.~Aidelsburger, M.~Lohse, C.~Schweizer, M.~Atala, J.~T. Barreiro,
  S.~Nascimbene, N.~Cooper, I.~Bloch and N.~Goldman,
\newblock \emph{{Measuring the Chern number of Hofstadter bands with ultracold
  bosonic atoms}},
\newblock Nature Physics \textbf{11}(2), 162 (2015),
\newblock \doi{/10.1038/nphys3171}.

\bibitem{stuhl2015visualizing}
B.~Stuhl, H.-I. Lu, L.~Aycock, D.~Genkina and I.~Spielman,
\newblock \emph{{Visualizing edge states with an atomic Bose gas in the quantum
  Hall regime}},
\newblock Science \textbf{349}(6255), 1514 (2015),
\newblock \doi{/10.1126/science.aaa8515}.

\bibitem{jane2003simulation}
E.~Jan{\'e}, G.~Vidal, W.~D{\"u}r, P.~Zoller and J.~Cirac,
\newblock \emph{Simulation of quantum dynamics with quantum optical systems},
\newblock Quantum Information and Computation \textbf{3}(1), 15 (2003),
\newblock \doi{10.26421/QIC3.1-2}.

\bibitem{blatt2012quantum}
R.~Blatt and C.~F. Roos,
\newblock \emph{Quantum simulations with trapped ions},
\newblock Nature Physics \textbf{8}(4), 277 (2012),
\newblock \doi{/10.1038/nphys2252}.

\bibitem{bose1924plancks}
Bose,
\newblock \emph{Plancks gesetz und lichtquantenhypothese},
\newblock Zeitschrift f{\"u}r Physik \textbf{26}(1), 178 (1924),
\newblock \doi{/10.1007/BF01327326}.

\bibitem{shi1998finite}
H.~Shi and A.~Griffin,
\newblock \emph{{Finite-temperature excitations in a dilute Bose-condensed
  gas}},
\newblock Physics Reports \textbf{304}(1-2), 1 (1998),
\newblock \doi{10.1016/S0370-1573(98)00015-5}.

\bibitem{andersen2004theory}
J.~O. Andersen,
\newblock \emph{{Theory of the weakly interacting Bose gas}},
\newblock Reviews of modern physics \textbf{76}(2), 599 (2004),
\newblock \doi{/10.1103/RevModPhys.76.599}.

\bibitem{Lee2016}
T.~E. Lee,
\newblock \emph{{Anomalous Edge State in a Non-Hermitian Lattice}},
\newblock Phys. Rev. Lett. \textbf{116}, 133903 (2016),
\newblock \doi{10.1103/PhysRevLett.116.133903}.

\bibitem{Kunst2018}
F.~K. Kunst, E.~Edvardsson, J.~C. Budich and E.~J. Bergholtz,
\newblock \emph{{Biorthogonal Bulk-Boundary Correspondence in Non-Hermitian
  Systems}},
\newblock Phys. Rev. Lett. \textbf{121}, 026808 (2018),
\newblock \doi{10.1103/PhysRevLett.121.026808}.

\bibitem{Yao2018}
S.~Yao, F.~Song and Z.~Wang,
\newblock \emph{{Non-Hermitian Chern Bands}},
\newblock Phys. Rev. Lett. \textbf{121}, 136802 (2018),
\newblock \doi{10.1103/PhysRevLett.121.136802}.

\bibitem{Gong2018}
Z.~Gong, Y.~Ashida, K.~Kawabata, K.~Takasan, S.~Higashikawa and M.~Ueda,
\newblock \emph{{Topological Phases of Non-Hermitian Systems}},
\newblock Phys. Rev. X \textbf{8}, 031079 (2018),
\newblock \doi{10.1103/PhysRevX.8.031079}.

\bibitem{El-Ganainy2018}
R.~El-Ganainy, K.~G. Makris, M.~Khajavikhan, Z.~H. Musslimani, S.~Rotter and
  D.~N. Christodoulides,
\newblock \emph{{Non-Hermitian physics and $\mathcal{PT}$ symmetry}},
\newblock Nature Physics \textbf{14}(1), 11 (2018),
\newblock \doi{/10.1038/nphys4323}.

\bibitem{Nakagawa2018}
M.~Nakagawa, N.~Kawakami and M.~Ueda,
\newblock \emph{{Non-Hermitian Kondo Effect in Ultracold Alkaline-Earth
  Atoms}},
\newblock Phys. Rev. Lett. \textbf{121}, 203001 (2018),
\newblock \doi{10.1103/PhysRevLett.121.203001}.

\bibitem{Shen2018}
H.~Shen and L.~Fu,
\newblock \emph{{Quantum Oscillation from In-Gap States and a Non-Hermitian
  Landau Level Problem}},
\newblock Phys. Rev. Lett. \textbf{121}, 026403 (2018),
\newblock \doi{10.1103/PhysRevLett.121.026403}.

\bibitem{Wu2019}
Y.~Wu, W.~Liu, J.~Geng, X.~Song, X.~Ye, C.-K. Duan, X.~Rong and J.~Du,
\newblock \emph{Observation of parity-time symmetry breaking in a single-spin
  system},
\newblock Science \textbf{364}(6443), 878 (2019),
\newblock \doi{/10.1126/science.aaw8205}.

\bibitem{Yamamoto2019}
K.~Yamamoto, M.~Nakagawa, K.~Adachi, K.~Takasan, M.~Ueda and N.~Kawakami,
\newblock \emph{{Theory of Non-Hermitian Fermionic Superfluidity with a
  Complex-Valued Interaction}},
\newblock Phys. Rev. Lett. \textbf{123}, 123601 (2019),
\newblock \doi{10.1103/PhysRevLett.123.123601}.

\bibitem{Song2019}
F.~Song, S.~Yao and Z.~Wang,
\newblock \emph{{Non-Hermitian Skin Effect and Chiral Damping in Open Quantum
  Systems}},
\newblock Phys. Rev. Lett. \textbf{123}, 170401 (2019),
\newblock \doi{10.1103/PhysRevLett.123.170401}.

\bibitem{Yang2019}
Z.~Yang and J.~Hu,
\newblock \emph{{Non-Hermitian Hopf-link exceptional line semimetals}},
\newblock Phys. Rev. B \textbf{99}, 081102 (2019),
\newblock \doi{10.1103/PhysRevB.99.081102}.

\bibitem{Hamazaki2019}
R.~Hamazaki, K.~Kawabata and M.~Ueda,
\newblock \emph{{Non-Hermitian Many-Body Localization}},
\newblock Phys. Rev. Lett. \textbf{123}, 090603 (2019),
\newblock \doi{10.1103/PhysRevLett.123.090603}.

\bibitem{Kawabata2019}
K.~Kawabata, T.~Bessho and M.~Sato,
\newblock \emph{{Classification of Exceptional Points and Non-Hermitian
  Topological Semimetals}},
\newblock Phys. Rev. Lett. \textbf{123}, 066405 (2019),
\newblock \doi{10.1103/PhysRevLett.123.066405}.

\bibitem{Kawabata2019a}
K.~Kawabata, S.~Higashikawa, Z.~Gong, Y.~Ashida and M.~Ueda,
\newblock \emph{{Topological unification of time-reversal and particle-hole
  symmetries in non-Hermitian physics}},
\newblock Nature Communications \textbf{10}(1), 297 (2019),
\newblock \doi{/10.1038/s41467-018-08254-y}.

\bibitem{Lee2019}
C.~H. Lee, L.~Li and J.~Gong,
\newblock \emph{{Hybrid Higher-Order Skin-Topological Modes in Nonreciprocal
  Systems}},
\newblock Phys. Rev. Lett. \textbf{123}, 016805 (2019),
\newblock \doi{10.1103/PhysRevLett.123.016805}.

\bibitem{Yokomizo2019}
K.~Yokomizo and S.~Murakami,
\newblock \emph{{Non-Bloch Band Theory of Non-Hermitian Systems}},
\newblock Phys. Rev. Lett. \textbf{123}, 066404 (2019),
\newblock \doi{10.1103/PhysRevLett.123.066404}.

\bibitem{Jin2020}
L.~Jin, H.~C. Wu, B.-B. Wei and Z.~Song,
\newblock \emph{{Hybrid exceptional point created from type-III Dirac point}},
\newblock Phys. Rev. B \textbf{101}, 045130 (2020),
\newblock \doi{10.1103/PhysRevB.101.045130}.

\bibitem{Lourenifmmodemboxccelseccfio2018}
J.~A.~S. Louren\ifmmode~\mbox{\c{c}}\else \c{c}\fi{}o, R.~L. Eneias and R.~G.
  Pereira,
\newblock \emph{{Kondo effect in a $\mathcal{PT}$-symmetric non-Hermitian
  Hamiltonian}},
\newblock Phys. Rev. B \textbf{98}, 085126 (2018),
\newblock \doi{10.1103/PhysRevB.98.085126}.

\bibitem{Mu2020}
S.~Mu, C.~H. Lee, L.~Li and J.~Gong,
\newblock \emph{{Emergent Fermi surface in a many-body non-Hermitian fermionic
  chain}},
\newblock Phys. Rev. B \textbf{102}, 081115 (2020),
\newblock \doi{10.1103/PhysRevB.102.081115}.

\bibitem{Okuma2019}
N.~Okuma and M.~Sato,
\newblock \emph{{Topological Phase Transition Driven by Infinitesimal
  Instability: Majorana Fermions in Non-Hermitian Spintronics}},
\newblock Phys. Rev. Lett. \textbf{123}, 097701 (2019),
\newblock \doi{10.1103/PhysRevLett.123.097701}.

\bibitem{Berry2004}
M.~V. Berry,
\newblock \emph{{Physics of Nonhermitian Degeneracies}},
\newblock Czechoslovak Journal of Physics \textbf{54}(10), 1039 (2004),
\newblock \doi{/10.1023/B:CJOP.0000044002.05657.04}.

\bibitem{Heiss2012}
W.~D. Heiss,
\newblock \emph{The physics of exceptional points},
\newblock Journal of Physics A: Mathematical and Theoretical \textbf{45}(44),
  444016 (2012),
\newblock \doi{/10.1088/1751-8113/45/44/444016}.

\bibitem{Miri2019}
M.-A. Miri and A.~Alù,
\newblock \emph{Exceptional points in optics and photonics},
\newblock Science \textbf{363}(6422), eaar7709 (2019),
\newblock \doi{/10.1126/science.aar7709}.

\bibitem{Zhang2020}
X.~Zhang and J.~Gong,
\newblock \emph{{Non-Hermitian Floquet topological phases: Exceptional points,
  coalescent edge modes, and the skin effect}},
\newblock Phys. Rev. B \textbf{101}, 045415 (2020),
\newblock \doi{10.1103/PhysRevB.101.045415}.

\bibitem{Doppler2016}
J.~Doppler, A.~A. Mailybaev, J.~Böhm, U.~Kuhl, A.~Girschik, F.~Libisch, T.~J.
  Milburn, P.~Rabl, N.~Moiseyev and S.~Rotter,
\newblock \emph{Dynamically encircling an exceptional point for asymmetric mode
  switching},
\newblock Nature \textbf{537}(7618), 76 (2016),
\newblock \doi{/10.1038/nature18605}.

\bibitem{Xu2016}
H.~Xu, D.~Mason, L.~Jiang and J.~G.~E. Harris,
\newblock \emph{Topological energy transfer in an optomechanical system with
  exceptional points},
\newblock Nature \textbf{537}(7618), 80 (2016),
\newblock \doi{/10.1038/nature18604}.

\bibitem{Assawaworrarit2017}
S.~Assawaworrarit, X.~Yu and S.~Fan,
\newblock \emph{Robust wireless power transfer using a nonlinear
  parity-time-symmetric circuit},
\newblock Nature \textbf{546}(7658), 387 (2017),
\newblock \doi{/10.1038/nature22404}.

\bibitem{Wiersig2014}
J.~Wiersig,
\newblock \emph{{Enhancing the Sensitivity of Frequency and Energy Splitting
  Detection by Using Exceptional Points: Application to Microcavity Sensors for
  Single-Particle Detection}},
\newblock Phys. Rev. Lett. \textbf{112}, 203901 (2014),
\newblock \doi{10.1103/PhysRevLett.112.203901}.

\bibitem{Wiersig2016}
J.~Wiersig,
\newblock \emph{{Sensors operating at exceptional points: General theory}},
\newblock Phys. Rev. A \textbf{93}, 033809 (2016),
\newblock \doi{10.1103/PhysRevA.93.033809}.

\bibitem{Hodaei2017}
H.~Hodaei, A.~U. Hassan, S.~Wittek, H.~Garcia-Gracia, R.~El-Ganainy, D.~N.
  Christodoulides and M.~Khajavikhan,
\newblock \emph{Enhanced sensitivity at higher-order exceptional points},
\newblock Nature \textbf{548}(7666), 187 (2017),
\newblock \doi{/10.1038/nature23280}.

\bibitem{Chen2017}
W.~Chen, {\c{S}}.~Kaya~{\"O}zdemir, G.~Zhao, J.~Wiersig and L.~Yang,
\newblock \emph{Exceptional points enhance sensing in an optical microcavity},
\newblock Nature \textbf{548}(7666), 192 (2017),
\newblock \doi{/10.1038/nature23281}.

\bibitem{yang1962concept}
C.~N. Yang,
\newblock \emph{{Concept of off-diagonal long-range order and the quantum
  phases of liquid He and of superconductors}},
\newblock Reviews of Modern Physics \textbf{34}(4), 694 (1962),
\newblock \doi{/10.1103/RevModPhys.34.694}.

\bibitem{jin2009solutions}
L.~Jin and Z.~Song,
\newblock \emph{{Solutions of $\mathcal{PT}$-symmetric tight-binding chain and
  its equivalent Hermitian counterpart}},
\newblock Physical Review A \textbf{80}(5), 052107 (2009),
\newblock \doi{/10.1103/PhysRevA.80.052107}.

\bibitem{zhang2013self}
X.~Zhang, L.~Jin and Z.~Song,
\newblock \emph{{Self-sustained emission in semi-infinite non-Hermitian systems
  at the exceptional point}},
\newblock Physical Review A \textbf{87}(4), 042118 (2013),
\newblock \doi{/10.1103/PhysRevA.87.042118}.

\bibitem{jin2010physics}
L.~Jin and Z.~Song,
\newblock \emph{{Physics counterpart of the $\mathcal{PT}$ non-Hermitian
  tight-binding chain}},
\newblock Physical Review A—Atomic, Molecular, and Optical Physics
  \textbf{81}(3), 032109 (2010),
\newblock \doi{10.1103/PhysRevA.81.032109}.

\bibitem{jin2011partitioning}
L.~Jin and Z.~Song,
\newblock \emph{{Partitioning technique for discrete quantum systems}},
\newblock Physical Review A—Atomic, Molecular, and Optical Physics
  \textbf{83}(6), 062118 (2011),
\newblock \doi{10.1103/PhysRevA.83.062118}.

\bibitem{jin2011physical}
L.~Jin and Z.~Song,
\newblock \emph{{A physical interpretation for the non-Hermitian Hamiltonian}},
\newblock Journal of Physics A: Mathematical and Theoretical \textbf{44}(37),
  375304 (2011),
\newblock \doi{10.1088/1751-8113/44/37/375304}.

\end{thebibliography}

%%%%%%%%%% END TODO: BIBLIOGRAPHY

\end{document}